\documentclass[notitlepage,superscriptaddress,twocolumn,showpacs,nobalancelastpage,aps,prl]{revtex4-1}
\usepackage[english]{babel}
\RequirePackage[T1]{fontenc}
\RequirePackage{times} 

\usepackage{comment}
\usepackage{amsfonts}
\usepackage{subfigure}
\usepackage{amsmath}
\usepackage{amssymb}
\usepackage{bm}
\usepackage{graphicx,epstopdf}
\usepackage{appendix}
\usepackage{array}
\usepackage{color}
\usepackage{mathrsfs}
\usepackage[hidelinks]{hyperref}
\usepackage{cancel}
\usepackage{my_symbols}
\usepackage[normalem]{ulem}
\usepackage{CJK}
\providecommand{\U}[1]{\protect\rule{.1in}{.1in}}

{\par}

\newcommand\rmv{\bgroup\markoverwith {\textcolor{red}{\rule[0.5ex]{2pt}{0.4pt}}}\ULon}

\begin{document}

\begin{CJK*}{UTF8}{gbsn} 
\title{Synchronized spin-photon coupling in a microwave cavity}
\author{Vahram L. Grigoryan}
\affiliation{The Center for Advanced Quantum Studies and Department of Physics, Beijing Normal University, Beijing 100875, China}
\affiliation{Beijing Normal University, Zhuhai, 519087,  Guangdong, China }
\author{Ka Shen}
\affiliation{The Center for Advanced Quantum Studies and Department of Physics, Beijing Normal University, Beijing 100875, China}
\author{Ke Xia}
\email[Corresponding author:~]{kexia@bnu.edu.cn}
\affiliation{The Center for Advanced Quantum Studies and Department of Physics, Beijing Normal University, Beijing 100875, China}
\affiliation{Synergetic Innovation Center for Quantum Effects and Applications (SICQEA), Hunan Normal University, Changsha 410081, China}
\begin{abstract}
We study spin-photon coupling in cavity in the presence of relative phase shift between two ferromagnetic resonance driving forces. We show that the anticrossing gap can be manipulated by varying the relative phase. Increasing the phase difference leads to narrowing the anticrossing gap of hybridized modes and eventually to phase locked coupling at the value of relative phase of $\pi.$ The FMR and cavity modes become phase locked and oscillate at the same frequency near the resonance frequency. 
Characteristic linewidth drop and transmission amplitude enhancement are demonstrated. The phase resolved spin-photon coupling can be used both for phase imaging and controlling coupling parameters.

\end{abstract}
\maketitle
\end{CJK*}


Strong interaction of light with matter in condensed matter systems paves way for exploring a wide range of different physical phenomena: starting from observation and even manipulation of matter by light in atomic scale and light manipulation to exploration of the polariton \cite{mills_1974} in order to develop quantum information technology. Achieving strong coupling due to cooperative phenomena of spin ensembles \cite{dicke_1954,tavis_1968} triggered strong interest in strong magnon-photon interactions between magnetic materials with low dissipation and high quality microwave cavities \cite{kubo_2010,amsuss_2011,schuster_2010,probst_2013,sandner_2012,huebl_2013,tabuchi_2014,zhang_2014,lambert_2016}. Coherent coupling between single spin and microwave cavity photons \cite{viennot_2015}, ferromagnetic magnon and a superconducting qubit \cite{tabuchi_2015} as well as cavity photons and magnons in magnetic materials \cite{huebl_2013,tabuchi_2014,zhang_2014,lambert_2016} have been reported. Indirect coupling between spins, mediated by cavity, have been achieved for cavity \cite{haroche_2006} and circuit quantum electrodynamics \cite{blais_2004,wallraff_2004}. In addition to widely used microwave transmission measurements of magnon-photon coupling at room temperature, electrical detection method has been recently demostrated by Hu's group \cite{bai_2015}. Theoretically, the spin-photon coupling has been formulated by means of scattering theory \cite{cao_2015} as well as simple semiclassical model \cite{bai_2015}.  The relevance of the classical picture to quantum mechanical picture has been discussed elsewhere \cite{harder_2016}. It was demonstrated that, although the coupling does not affect the intrinsic Gilbert damping, the FMR linewidth ($\Delta H$) always increases \cite{bai_2015} when FMR frequency approaches to resonance.

\begin{figure}[t]
  \begin{tabular}{c}
\includegraphics[width=\columnwidth]{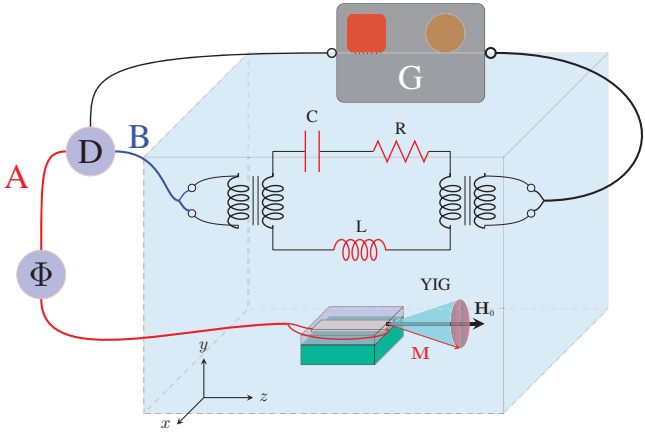}
\end{tabular}
\caption{(Color online) Schematic picture of the system. The microwave signal is generated in "G". The generated signal is divided into "A" and "B" in rf power divider "D". The signal "A" travels through phase shifter "$\Phi$," to shorted strip line. The other signal travels directly to the cavity through path "B". The microwave in the cavity is modeled by LCR circuit. 
}
\label{fig:schematic}
\end{figure}

To overcome the drawback of the linewidth broadening due to coupling induced extrinsic damping \cite{bai_2015},  while having strong coupling, we consider spin-photon coupling in cavity resonator when, together with the magnetic component of microwave field in the cavity, an additional local FMR driving force exists with a relative phase shift ($\Phi$) between them. 
 In \Figure{fig:schematic} we show the schematic picture of the system under study. In this device the microwave signal from a broadband microwave generator "G" is directed via a coaxial cable to a rf power divider "D" \cite{wirthmann_2010,cao_2013}, which coherently splits the microwave into two different beams. One of the beams then travels through a microwave phase shifter \cite{wirthmann_2010,cao_2013} "$\Phi$" by path "A" to an integrated \cite{patent} strip line on an insulator nonmagnetic layer. The YIG film is on top of the strip line. 
The magnetic field, created by microwave current from path "A"  acts locally on ferromagnetic insulator magnetization as an additional FMR driving force \cite{twofm1,gui_2013}. The other beam remains undisturbed and travels by path "B" through coaxial cable to microwave cavity resonator (the blue box in \Figure{fig:schematic}).  
Thus, the YIG magnetization effectively feels two time dependent magnetic fields, 
a local magnetic field $\bh^\ssf{A}$ and the magnetic component of the microwave inside the cavity $\bh^\ssf{B}$. 
 We assume that the magnitude $h^\ssf{A}=\delta h^{\ssf{B}},$ where $\delta$ can be controlled by the divider. 
 We show that for the coupling 
the relative phase ($\Phi$) between two paths plays the leading role in the FMR line shape. The spectrum of hybridized modes (polaritons) depends on the relative phase $\Phi$. Particularly, the gap between two  polariton modes, close to the resonance frequencies, can be tuned by varying $\Phi$. More interestingly, phase locked coupling regime can be achieved by tuning the relative phase to $\pi.$ Important consequence of the phase locked coupling is that together with achievement of strongly coupled modes, the FMR linewidth, at the same time, becomes very narrow with enhanced output microwave power. This phase resolved spin  photon coupling can be used both for phase imaging and controlling coupling parameters.

Simple semiclassical picture describing spin-phonon interaction in the cavity is based on the combination of a microwave LCR and Landau-Lifshitz-Gilbert (LLG) equations \cite{bloembergen_1954,bai_2015}. The coupling of the magnetization dynamics with microwave is established via two classical coupling mechanisms. First one is known as the Faraday induction of FMR \cite{silva_1999} which induces voltage in LCR circuit due to the precessing magnetization. The other coupling mechanism is governed by Ampere's law producing magnetic field (that is the magnetic component of microwave).

We consider a ferromagnetic insulator lying in $\hxx$-$\hzz$ plane with an in-plane magnetic easy axis pointing in $\hzz$ direction due to crystal anisotropy, dipolar and external magnetic field. The LCR circuit in the picture illustrates the theoretical model of the electromagnetic field in the cavity.
 The RLC circuit equation of two crossed coils parallel to the $\hxx$ and $\hyy$ directions, in which the microwave current $\bj^\ssf{B}(t)$ is driven by the RF voltage is
\begin{align}
&L \dot{\bj^\ssf{B}} +R \bj^\ssf{B} +\smlb{1/ C}\int \bj^\ssf{B} dt=\bV^F \label{eq:LCR},
\end{align}
where $L,$ $C$, and $R$ are inductor, capacitor, and resistor, respectively. The driving voltage 
 $\bV^F$ is induced by precessing magnetization according to Faraday induction
\begin{align}
&V^F_x\smlb{t}=K_c L\dot{m}_y,~V^F_y\smlb{t}=-K_c L\dot{m}_x\label{eq:2},
\end{align}
 The magnetization precession in the magnetic sample is governed by the LLG equation \cite{gilbert_2004}
\begin{align}
& \dot{\mb}=\mb\times\gamma\bH  - \alpha \mb\times \dot{\mb}\label{eq:LLG},
\end{align}
where $\mb=\bM/M_{s}$ is the magnetization direction in FM with $M_{s}$ being the saturation magnetization. $\alpha$ is the intrinsic Gilbert damping parameter.  $\bH=\bH_{0}+\bh^\ssf{A}+\bh^\ssf{B}$ is the effective magnetic field in FM with $\bH_0=H_0\hzz$ being the the sum of external magnetic, anisotropy, and dipolar fields aligned with $\hzz$ direction. $\bh^\ssf{B}=\bh e^{-i\omega t}$  and $\bh^\ssf{A}= \delta\bh^\ssf{B} e^{-i\Phi}$ are the magnetic fields from path "B" and "A," respectively. $\Phi$ is the phase shift between them and $\delta$ is the ratio between their amplitudes, which can be controled by experiment \cite{wirthmann_2010,cao_2013}.
 Using the form of magnetization $\mb =\hzz+\mb_\perp e^{-i\omega t}$ the linearized LLG equation becomes
\begin{equation}
m^+ \smlb{\omega-\omega_r+i\alpha \omega}+\smlb{1+\delta e^{i\Phi} } \omega_m h^+=0\label{eq:LLG1},
\end{equation}
where  $m^+=m_{x}+i m_{y}$ is in-plane component of the magnetization, $\omega_m=\gamma M_s$ and $\omega_r \simeq \gamma H_0$  with $\gamma$ being the gyromagnetic ratio, $h^+=h_{x}+ih_{y}.$ 
 The linearized equation of motion in \Eq{eq:LCR} solved by the form 
  $\bj^\ssf{B}=\bj_\perp^\ssf{B} e^{i\omega t}$ leads to

  \begin{figure}[t!]
\includegraphics[width=.9\columnwidth]{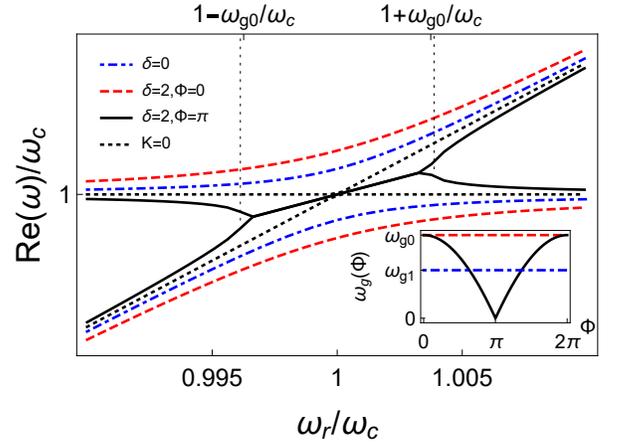}
\caption{(Color online) Dispersion relation of polariton modes, i.e., FMR mode coupled to microwave mode for different values of $\delta$ and relative phase $\Phi$. The inset shows the gap between two polariton modes, $\omega_g,$ as a function of the phase shift at $\omega_r=\omega_c.$
}
\label{fig:spectrum}
\end{figure}
\begin{figure*}[t]
  \begin{tabular}{ccc}
\includegraphics[width=.66\columnwidth]{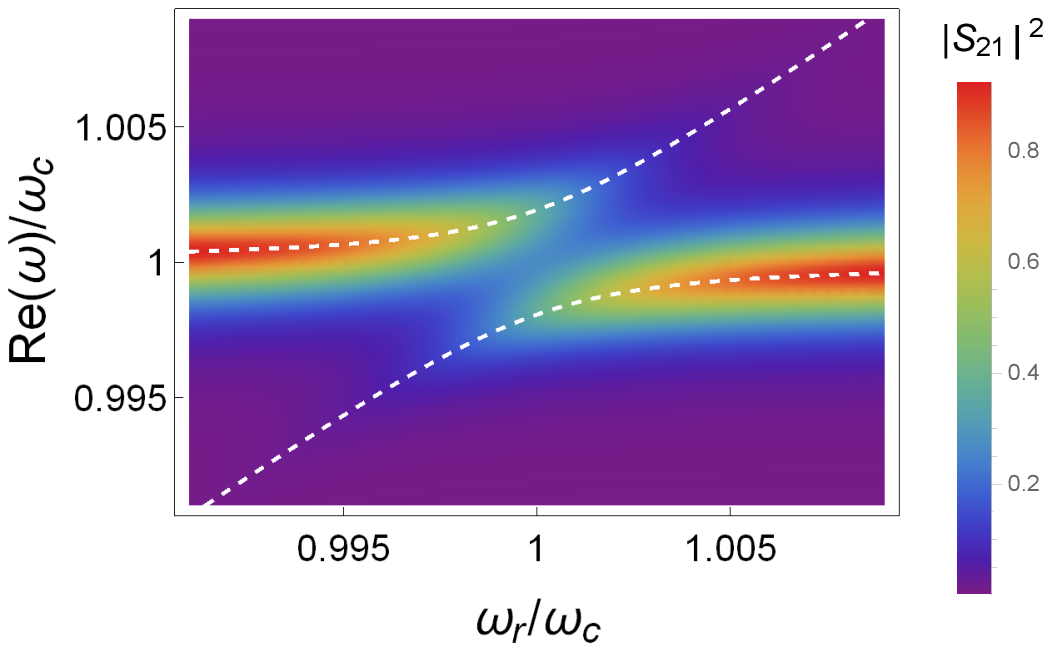}&
\includegraphics[width=.66\columnwidth]{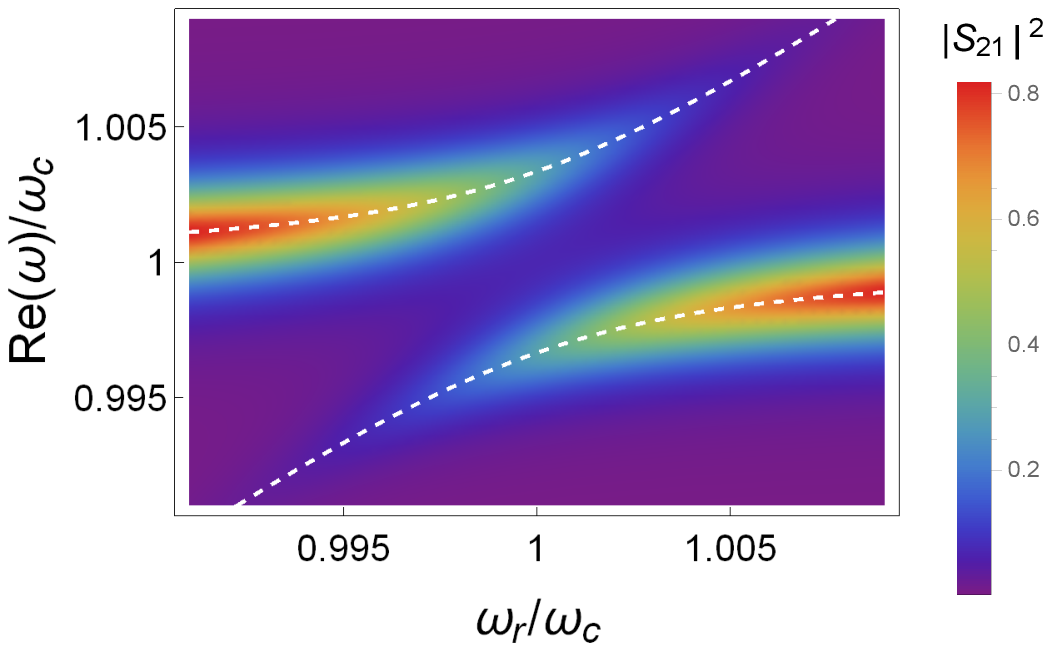}&
\includegraphics[width=.66\columnwidth]{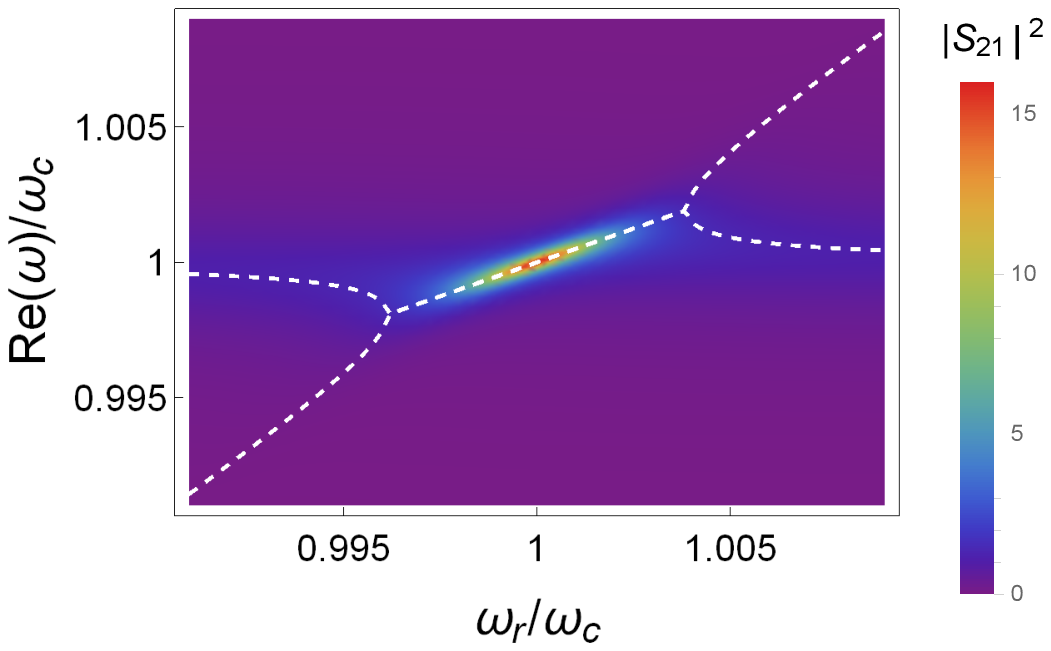}
\end{tabular}
\caption{(Color online) Dashed line stand for dispersion relation of coupled modes. The colored area shows the transmission amplitude defined in \Eq{eq:transmission} for (a) $\delta=0,$ (b) $\delta=2,$ $\Phi=0,$ and (c) $\delta=2,$ $\Phi=\pi,.$
}
\label{fig:spectrum1}
\end{figure*}

 \begin{align}
&\Omega  \smatrix{ m^+\\ h^+}=0\qwith\nn
 & \Omega \equiv\smatrix{ \omega-\omega_r + i\alpha \omega  & \omega_{m} \smlb{1+\delta e^{i\Phi}}  \\ \omega^2 K ^2 & {\omega^2+ 2i \beta \omega \omega_c - \omega_c^2 }} \label{eq:mat}
  \end{align}
  where due to the Ampere's law
\begin{align}
&h_{x}^\ssf{B}=K_m j_y^\ssf{B},~h_{y}^\ssf{B}=-K_m j_x^\ssf{B} \label{eq:3}
\end{align}
which places a torque on the FM magnetization. Here $j_{x,y}^\ssf{B}$ are the current components in the circuit, satisfying \Eq{eq:LCR}. Parameters $K_c$ and $K_m$ are coupling parameters. $K\simeq \sqrt{K_cK_m},$ the cavity frequency is $\omega_c=1/\sqrt{LC},$ and $\beta=R/\smlb{2L\omega_c}$ is the cavity mode damping. By solving the complex eigenfrequencies of \Eq{eq:mat} ($\det\Omega=0$) at given magnetic field we obtain roots for $\omega$. The two positive real components of $\omega$ determine resonant frequencies, while imaginary parts describe damping of the coupled system.

In \Figure{fig:spectrum} we plot the dispersion spectrum $\re{\omega(\omega_r)}$ (normalized by $\omega_c$) for different values of relative phase $\Phi.$ Dotted lines correspond to cavity mode and the Kittel's mode in the absence of spin-photon coupling. 
We set equal damping for FMR and LCR as $\alpha=\beta=0.002,$ $\omega_m=\gamma M_s\approx 0.075 \omega_c,$ the coupling parameter is $K=0.01$ \cite{linhui_2015_1}, with $\omega_c/2\pi\approx 10.5$ GHz.

We further discuss different values for free parameters $\delta$ and $\Phi.$ First, the blue dash-dotted lines in \Figure{fig:spectrum} is the spectrum (normalized by $\omega_c$) in case when no second path exist, which means $\delta=0.$ This corresponds to usual spin-photon coupling with the characteristic anticrossing of two modes \cite{bai_2015}. Here, the gap between two modes at FMR resonance ($\omega_r=\omega_c$) is proportional to the coupling constant $K$ and in strong coupling regime ($K>\alpha,\beta$) can be approximated as $\omega_{g1}\equiv\omega_{g}\smlb{\delta=0}=K\sqrt{2\omega_m \omega_c}$ \cite{linhui_2015_1}. Beyond the widely studied regime, we distinguish two different sets of parameters $\delta$ and $\Phi,$ (\textit{i}) $\Phi=0,~\delta=2$ and (\textit{ii}) $\Phi=\pi,~\delta=2.$ 
 For the former case no phase shift is introduced between two paths, thus two FMR driving forces are in phase. In \Figure{fig:spectrum} the red dashed curve shows the spectrum in the case (\textit{i}), and the solid black curve is for (\textit{ii}). In the case where no phase shift is introduced we can see that the coupling increases with increasing $\delta.$ Moreover, as shown in the inset of \Figure{fig:spectrum}, tuning the relative phase for $\delta=2$ changes the gap between two modes at FMR resonance in the range of the coupling bandwidth (the range of FMR frequencies, projected by dotted lines in the figure, where the polariton modes occur), equal to $2\omega_{g0}/\omega_c,$ where $\omega_{g0}\equiv\omega_{g}\smlb{\delta=2,\Phi=0}=K\sqrt{6\omega_m \omega_c}$. The relative phase shift causes principally different picture of spin-photon coupling spectrum as well as the transmission amplitude. In the spectra, for $\Phi=0$ at $\omega_r<\omega_c$ the frequency of FMR mode decreases with increasing frequency of cavity mode and the opposite at $\omega_r>\omega_c.$ The picture is now inverse at $\Phi=\pi$: due to the phase shift, the FMR mode frequency increases at $\omega_r<\omega_c$ together with decreasing of cavity mode frequency and the opposite at $\omega_r>\omega_c$. As a consequence a gap is opened due to the coupling in $\Phi=0$ case and for $0 \leq \Phi \leq \pi$ it decreases with increasing $\Phi$. The frequencies of two modes get "pulled" \cite{slavin_2009} toward each other (see the inset of \Figure{fig:spectrum}) and eventually leads to phase locked synchronization of two modes at $\Phi = \pi$ when two modes start to share the same frequency $\smlb{\omega_c+\omega_r}/2.$ Notice that although the studies of synchronization usally refer to non-linear system \cite{rippard_2005,slavin_2009}, the synchornization has also been demonstrated in purely linear systems \cite{heinrich_2003,tserkovnyak_2005,zhang_2017,harder_2017,bernier_2018}. Our study here thus provide a new version of synchronization in linear system. In the presence of phase shift the dependence of gap on $\Phi$ at $\omega_r=\omega_c$ can be approximated as $\omega_{g}\smlb{\Phi}=\sqrt{3} \omega_{g1} \abs{\cos\smlb{\Phi/2}}.$ 

 The physical picture of phase locked coupling in the present system is the following: a localized magnetic moment "feels" the effective magnetic field from two microwave sources and changes its direction causing a voltage in LCR circuit due to the Faraday induction. This voltage causes a current in the circuit, which, in turn, produces magnetic field due to Ampere's law and exerts torque on the magnetization.   
 When there is no phase shift between $\bh^\ssf{A}$ and $\bh^\ssf{B}$ ($\Phi=0$), 
 energy exchange takes place between microwave in the cavity and oscillating magnetization, where one drives the other. 
In the case of two driving forces with opposite phase ($\Phi=\pi$), 
the energy loss due to damping of the magnetization precession is compensated by the torque coming from $\bh^\ssf{A}.$   
As a result, instead of performing as a dissipation channel in usual case \cite{bai_2015}, the ferromagnetic insulator under $\Phi=\pi$ absorbs energy from $\bh^\ssf{A}$ field and behaves as a pumping source of $\bh^\ssf{B}$ field , leading to an enhancement of $\bh^\ssf{B}$ field.

\begin{figure}[t]
  \begin{tabular}{c}
\includegraphics[width=\columnwidth]{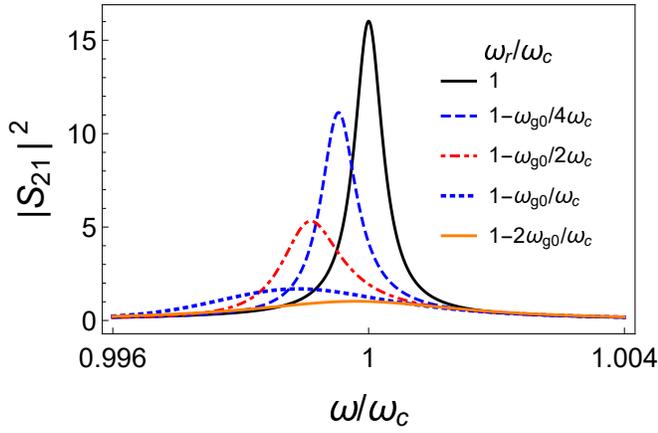}
\end{tabular}
\caption{(Color online) The transmission amplitude in phase locked regime as a function of $\omega$ at different values of FMR frequency.
}
\label{fig:spectrum2}
\end{figure}

To study the signal power enhancement, we calculate the transmission amplitude using input-output formalism from \Eq{eq:mat} \cite{bai_2015,twofm,twofm1}
\begin{align}
&\Omega  \smatrix{ m^+\\ h^+}=\smatrix{0\\ \omega ^2 h_{0}^+} \qand \nn
&S_{21}=\Gamma h^+ / h_{0}^+=\Gamma {\omega^2\smlb{\omega +i\alpha \omega -\omega_r}\ov \det{\Omega}}
\label{eq:transmission},
\end{align}
where $h_{0}^+$ is the input magnetic field driving the system $\Gamma$ characterizes the cavity/cable impedance mismatch \cite{bai_2015}. In \Figure{fig:spectrum1} we plot the spectrum of the coupled system, where the dashed line is $Re\smlb{\omega\smlb{\omega_r}}$ and the colored area represents the transmission amplitude. In \Figure{fig:spectrum1} (a) we show the transmission amplitude, when $\delta=0.$ There we can see the usual anticrossing feature in the strong coupling regime. \Figure{fig:spectrum1} (b) shows the increased coupling when $\delta=2$ but no phase shift exists ($\phi=0$). The phase locking is shown in \Figure{fig:spectrum1} (c). It is seen that the transmission amplitude - proportional to transmission power - is lower at FMR frequencies far from resonance and increases dramatically when $\omega_r$ approaches to $\omega_c.$ \Figure{fig:spectrum2} shows the transmission amplitude evolution as a function of $\omega$ at different fixed values of FMR frequency $\omega_r.$ At FMR frequencies far from resonance, $\abs{\omega_r-\omega_c}\gg \omega_{g0}$, the transmission amplitude is not affected by the FMR and the transmission is due to the cavity mode and equal to that of empty cavity. Tuning the FMR frequency close to the resonance with the cavity mode, $\abs{\omega_r-\omega_c}\lesssim \omega_{g0},$ the transmission amplitude increases and reaches its maximum (black solid curve in \Figure{fig:spectrum2}) at resonance $\omega_r=\omega_c.$

\begin{figure}[t]
  \begin{tabular}{c}
\includegraphics[width=\columnwidth]{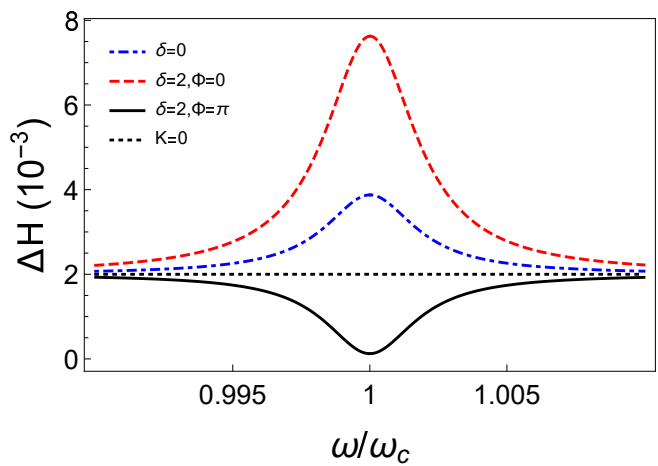}
\end{tabular}
\caption{(Color online) Normalized FMR linewidth $\Delta H$ as a function of $\omega.$
}
\label{fig:spectrum3}
\end{figure}

To analyze the second characteristic feature of phase locked coupling, namely the FMR linewidth drop, we solve \Eq{eq:mat} to obtain $\omega_r$ as a function of $\omega.$ In contrast to the solution for $\omega,$ here we have only one solution. The real part of the $\omega_r\smlb{\omega}$ is the FMR spectrum while the imaginary part is the linewidth $\Delta H=Im\smlb{\omega_r\smlb{\omega}} / Re \smlb{\omega_r\smlb{\omega}}.$ It has been demonstrated, that in the coupled cavity FMR system, the normalized linewidth is being changed by the coupling \cite{bai_2015}. In \Figure{fig:spectrum3} we plot the normalized linewidth dependence on $\omega.$ Dotted curve shows that in the absence of spin-photon coupling the linewidth is constant and determined by the damping constant $\alpha$. In the presence of spin-photon coupling but absence of beam "A" ($\delta=0$) the normalized linewidth is being increased near the resonance by the coupling (blue dot-dashed line in \Figure{fig:spectrum3}) \cite{bai_2015} and the increment is even larger for (\textit{i}) ($\Phi=0,$ $\delta=2$), shown by red dashed curve. In the case when a relative phase shift $\Phi = \pi$ is introduced the linewidth decreases (black solid curve in \Figure{fig:spectrum3}) near the resonance frequency as expected for the phase locked coupling.

In contrast to $\delta=0$ discussed in Ref. \onlinecite{bai_2015}, where the linewidth always increases as the FMR approaches the resonant coupling condition, no matter whether $\alpha < \beta$ or $\alpha > \beta,$ here both transmission coefficient and FMR linewidth depend on interplay between damping of two oscillators and coupling strength. At resonance ($\omega=\omega_c$), the linewidth dependence on damping coefficients and the phase $\Phi$ can be expressed as
\begin{equation}
\Delta H=\frac{2 \alpha  \beta +K^2 \smlb{\omega _M/\omega_c} (\delta  \cos (\Phi )+1)}{2 \beta -\delta  K^2 \smlb{\omega _M/\omega_c} \sin (\Phi )} \label{eq:damp},
\end{equation}
which reduces to
\begin{equation}
\Delta H=\alpha +(1-\delta  )\frac{K^2 \smlb{\omega _M/\omega_c} }{2 \beta } \label{eq:damp1}
\end{equation}
at $\Phi=\pi.$ It turns out that for $\delta<1$ the FMR linewidth increases in the presence of spin-photon coupling due to the coupling induced damping enhancement \cite{bai_2015}. However, the second driving force with $\pi$ phase shift ($\delta>1$) acts as an anti-damping field-like torque, which reduces the intrinsic damping and coupling-induced linewidth broadening. For a very small damping and strong coupling,  instability is reached when the effective linewidth becomes negative due to the phase-shifted driving force. 
In this case, the trajectory of the magnetization  can become very complicated and could possibly lead to a magnetization reversal to a different (meta)stable configuration. 
We believe that more systematic study of complicated dynamic phases in this regime would be useful by using numerical methods. We define a condition of stability of the system in such a way that the FMR linewidth is positive meaning that the phase shifted driving force can fully compensate the intrinsic and coupling induced damping. 
  At resonant frequency, where $\omega=\omega_c,$ this condition leads to (for $\delta=2$) $\alpha \beta=K^2 \omega_m/2\omega_c.$ Thus, large cavity damping can be compensated by low Gilbert damping.  Although the calculations here are made with equal damping parameters, the synchronization behavior is the same for other damping parameters satisfying $\Delta H>0$ condition, where $\Delta H$ is defined in \Eq{eq:damp} at resonant FMR frequency.

In summary, we study the spin-photon coupling in cavity in the presence of two FMR driving forces with a relative phase shift between them. We show that the anticrossing gap between the hybridized cavity and FMR modes can be controlled by a relative phase. Increasing the phase shift leads to reduction of the gap and eventually forming the phase locked spectrum when the relative phase equals to $\pi.$ At this regime two modes start to oscillate at the same frequency in $\abs{\omega_r-\omega_c}\lesssim \omega_{g,0}$ range. Linewidth drop and power enhancement are demonstrated.

A technical challenge for the experimental realization of the proposed synchronized coupling is 
to isolate the local field $\bh^\ssf{A}$ from the cavity, so that it will not introduce any unexpected influence on the cavity. 
This might be possible if the YIG film is capped by microwave absorbing and metallic shielding layers \cite{qin_2012}. 
We note, that in the proposed setup the cavity quality could be severely reduced when it is loaded by a stripline. However, the cavity damping change due to reduced quality factor is not essential and the predicted phenomenon still survives as long as $\Delta H>0$ is satisfied. 
Another option is to use an open cavity  \cite{huebl_2013}, where  one path travels through a phase shifter to a waveguide attached to a ferromagnetic insulator film \cite{zhu_2011}, while the other goes to a horn antenna, exposed to the YIG film. 
In that case the effect of the magnetic field from the horn antenna on the waveguide mode can be shielded by covering part (enclosing the waveguide overlapping area in the opposite side of YIG) of the YIG film.

We thank M. Weides and I. Boventer for discussion about experimental setup. This work was financially supported by National Key Research and Development Program of China (Grant No. 2017YFA0303300) and the National Natural Science Foundation of China (No.61774017, No. 11734004, and No. 21421003). K.S. acknowledges the Recruitment Program of Global Youth Experts.

\bibliographystyle{apsrev}

\appendix

\end{document}